\documentclass{llncs} 
\usepackage[
backend=biber,
	style=numeric,
	giveninits=true,
	safeinputenc,%required for arxiv
]{biblatex}
\usepackage[
	a4paper,top=3cm,bottom=2cm,left=3cm,
	right=3cm,marginparwidth=1.75cm,headheight=24pt]
{geometry}

\PassOptionsToPackage{dvipsnames}{xcolor}
\RequirePackage{comgipp}

\DeclareBibliographyCategory{preprintref}
\newcommand{\setPreprintRef}[1]{%
\def\theReferenceText{\fullcite{#1}}\addtocategory{preprintref}{#1}}

\setPreprintRef{CohlGS18}
\usepackage[english]{babel}
\usepackage[utf8]{inputenc}
\usepackage[T1]{fontenc}
\usepackage{latexml}
\usepackage{amsmath}
\usepackage{amsfonts}
\usepackage{graphicx}
\usepackage{url}
\usepackage{breakurl}
\usepackage[colorlinks=true, allcolors=blue,breaklinks]{hyperref}

\newcommand\Maple{{\sf MAPLE}}
\newcommand\TT{\rule{0pt}{2.2ex}}
\newcommand\TB{\rule[-0.9ex]{0pt}{0pt}}

\title{\vspace*{2cm}Automated Symbolic and Numerical Testing of\\DLMF Formulae using Computer Algebra Systems}
\author{\hspace{-0.8cm}Howard S.~Cohl,\inst{1}
Andr\'{e} Greiner-Petter,\inst{2} and Moritz Schubotz\inst{2}
}

\institute{\hspace{-0.15cm}Applied and Computational Mathematics Division,
National Institute of Standards and Technology,
Mission Viejo, CA, U.S.A.,
\email{howard.cohl@nist.gov}
\and%
\hspace{-0.10cm}Dept.~of Computer and Information Science,
University of Konstanz, Konstanz, Germany,
\email{first.last@uni-konstanz.de}
}

\begin{document}

\maketitle
\thispagestyle{firstpagestyle}

\begin{abstract}
\noindent We have developed an automated procedure for symbolic and numerical testing
of formulae extracted from the NIST Digital Library of Mathematical Functions
(DLMF).  For the NIST Digital Repository of Mathematical Formulae, we have
developed conversion tools from semantic \LaTeX\ to the Computer Algebra System
(CAS) \Maple\ which relies on Youssef's part-of-math tagger.  We convert a test data subset
of 4,078 semantic \LaTeX\ DLMF formulae
%extracted from the DLMF
to the native CAS representation and then apply an automated scheme for symbolic and numerical
testing and verification.  Our framework is implemented using Java and \Maple.
We describe in detail the conversion process which is required so that the
CAS can correctly interpret the mathematical representation of the formulae.
We describe the improvement of the effectiveness of our automated scheme through
incremental enhancement (making more precise) of the mathematical semantic markup
for the formulae.
\end{abstract}

%------------------------------------------------------------------------------
\vspace{-0.50cm}
\section{Problem and Current State}
\vspace{-0.0cm}

The NIST Digital Library of Mathematical Functions \cite{NIST:DLMF} (DLMF) is an online
digital mathematics library which focuses on special functions and orthogonal polynomials.
The DLMF is special in that it has been written like a book, but published online as a
digital resource. The DLMF contains a table of contents which links to specific chapters
(See Table \ref{firsttable1} for a summary of the 36 DLMF Chapters)
each of which has been written and edited by mathematicians who are experts in their specific
field of focus. Furthermore, the DLMF is constantly being maintained (edited, corrected,
updated) by a team of mathematicians and computer scientists who are determined to maintain
and preserve the high standard of mathematical accuracy and exposition.

The DLMF has been developed in such a way that mathematical metadata is constantly being
improved and restructured. For the organizational structure, semantic metadata has been
incorporated at the document, chapter, section, subsection, and paragraph levels. An effort
has been made to make more precise the nature of the objects which appear in the DLMF.
The DLMF's semantic realization of {\it mathematical} content has very much in common
with the way mathematics is written using \LaTeX, for the preparation of journal publications
and actual physical books. In fact, for the construction of the DLMF, Bruce Miller has developed
and actively maintains \LaTeXML\ \cite{LaTeXML1}, a program which is able to process a certain
flavor of \LaTeX\ and from it, generate a fully functional web site. We refer to this flavor of
\LaTeX\ as semantic \LaTeX. The website is then generated using \LaTeXML, from a collection of
semantic \LaTeX\ source documents, and from other documents including 
Cascading Style Sheets (CSS), 
WebGL (Web Graphics Library) controls, 
Mathematical Markup Language (\MathML),
\LaTeX\ style files, 
\LaTeXML\ binding files (\LaTeXML\ analogue of a style or class file),
Java, 
JavaScript, 
etc.~(see for instance \cite{MillerYoussef2003}).
The outcome is a highly sophisticated digital platform for accessing mathematical content on
the web.

One main difference between \LaTeX\ and semantic \LaTeX{} is the latter's ability to encapsulate mathematical
knowledge in such a way that it can then be extracted by a computer program, namely \LaTeXML{}, to
generate correctly formatted \MathML{} (see \cite{SchubotzGreiner18} for further information about this).
\LaTeXML{} is then able to interpret the content written in semantic \LaTeX\ in order to generate
presentation (and content) \MathML. This is then used to construct the DLMF web site and to,
%M: is that correct "to, for instance, display"
for instance, display metadata associated with the mathematical content. The mathematical semantic
content of the DLMF includes mathematics which appears within
the main body of the text, formulae (in various formulae environments), tables (in which
mathematics is displayed in a tabular environment), figures (in which mathematics is visualized),
operators, functions/symbols, variables (see \cite{POMTagger} for a nice description), etc.

Even though there certainly exist methods to convert \LaTeX\ expressions into 
CAS formats, these in practice are only really effective if one is dealing with 
elementary functions defined in terms of compositions of powers-laws, sums, 
differences, products, quotients, trigonometric/hyperbolic functions, 
and exponential/logarithmic functions. If one is dealing with more complicated 
special functions like Bessel functions, hypergeometric functions, Airy functions, 
elliptic integrals, elliptic functions, etc., then effective conversions from 
\LaTeX\ to CAS really do not exist, and our implementation described below is 
the first to be able to handle situations such as these. Also, several differences 
in CAS vs.~DLMF implementations must be captured in any effective translation,
including:~(1) the usage of $m=k^2$ for elliptic integrals/functions; (2) branch 
cuts for Legendre/Ferrers functions (and other functions); (3) and differences 
in normalizations of special functions. If one is unable to capture these subtleties
in a translation process for the DLMF, which require careful and detailed 
mathematical knowledge and implementation, then the translation will fail.

In a previous paper \cite{CohlDRMF3}, we described the conversion process that 
we have developed to input semantic \LaTeX\ and output a 
corresponding \Maple\footnote{The mention of 
specific products, trademarks, or brand
names is for purposes of identification only. Such mention is not to be interpreted 
in any way as an endorsement or certification of such products or brands by the 
National Institute of Standards and Technology, nor does it imply that the products 
so identified are necessarily the best available for the purpose. All trademarks 
mentioned herein belong to their respective owners.} CAS representation. Not 
surprisingly, this
conversion is continually in development. New ideas are being implemented in order to increase
the semantic enhancement of the original source (see for instance \cite{VMEXT}), as well as to
improve the effectiveness of our conversion software. In order to do this, we are also contributing
in order to assist in the development of the part-of-math (POM) tagger \cite{POMTagger}, which our conversion
program relies upon.

In this paper, we describe the process of extracting mathematical formulae 
from the DLMF, and then take advantage of the powerful mathematical semantic 
coverage of semantic \LaTeX\ used in the DLMF to convert to CAS representations 
whose sole purpose is to drive an automated scheme for verification of DLMF 
formulae.  In order to improve the precision of our DLMF formulae testing
reported on in \cite{CohlDRMF3}, we did the following:~(1) we updated our 
test formulae dataset; (2) fixed numerous translation bugs; (3) implemented a 
new method for handling constraints; and (4) added configuration files which 
make it easier to change the setup of our symbolic and numerical testing. 

\vspace{-0.30cm}
\section{Extraction of DLMF formulae}\label{sec:LatexToCAS}
\vspace{-0.0cm}

\subsection{First extraction scan}
\label{firstextraction}

In the first extraction scan, we extract mathematical formulae from 36~\LaTeX\ source files
which summarize the content of the 36 DLMF chapters 
formulae are extracted out of formula environments, which are (some custom)

\verb|{equation}|, \verb|{equationmix}|, \verb|{equationgroup}|, \verb|{align}|.
Each DLMF formula in the output text file is then represented as a single string on each line in semantic \LaTeX. In order to accomplish this, all lines of the source \LaTeX\
file are merged, except that the following character strings are removed: comments; space (and other) formatting
commands (such as \verb|\,| (used to insert a small horizontal space),
\verb|\!| (used to remove horizontal space), \verb|\\[..]| (used to introduce line breaks),
\verb|&| (used for alignment purposes),
\verb|\*| (used to force line breaking on multiplication), etc.); \verb|\MarkNotation|, \verb|\origref|,
\verb|\note|, \verb|\lxRefDeclaration|, \verb|\index|, \verb|\source|, \verb|\authorproof|

Along with the semantic \LaTeX\ source for the formula, we also extract two pieces of
metadata associated with that formula, a \verb|\constraint| environment, and also a \verb|\label|
environment. The \verb|\constraint| and \verb|\label| environments contain any constraints
and \LaTeX\ labels which have been directly associated with the formula within the semantic \LaTeX\
formula environment in which it was extracted.  If more than one formula is associated with
a \verb|\label| environment, we inherit that \verb|\label| to all of those formulae.

There are also certain \LaTeX\ commands or replacement macros which are specifically defined in
the preamble of certain chapter source files. These are replaced in order for our translator
to be able to correctly understand the corresponding derived semantic \LaTeX\ (e.g., in Chapters
\verb|AI|, \verb|CH| the entity \verb|\gamma| is replaced 
with the macro \verb|\EulerConstant| \cite[(5.2.3)]{NIST:DLMF}\footnote{The macro {\tt \textbackslash 
EulerConstant} is just one of many (currently unpublished) semantic DLMF \LaTeX\ semantic macros which have 
been developed by Bruce Miller for utilization in the DLMF. In fact, the macro names used in this 
manuscript, have been more recently updated. However, the macro set we are utilizing, is for those
in usage as of 9/16/2016.};
in Chapters \verb|JA|, \verb|MA|, \verb|LA|, \verb|EL|, the entities \verb|K| and
\verb|K|
\hspace{-0.10cm}\textquotesingle\hspace{0.03cm}
are replaced with the
semantic macros \verb|\CompEllIntKk@@{k}| and \verb|\CompEllIntCK@@{k}| respectively\footnote{These semantic macros
represent the complete and complementary elliptic integrals of the 
first kind \cite[(19.2.8-9)]{NIST:DLMF}}, etc.).
See Table \ref{firsttable1} for a description of the DLMF chapter codes.
An attempt is made to remove punctuation marks at the end of each formula since these are really
not a part of the formula itself.  One final time consuming task is the conversion, wherever required,
for the replacement of the entities \verb|i|, \verb|e|, and the \LaTeX\ command \verb|\pi| to the
semantic macros \verb|\iunit|, \verb|\expe|, and \verb|\cpi|\footnote{These semantic macros represent the mathematical
constants $i$, $e$ and $\pi$ \cite[(1.9.1), (4.2.11), (3.12.1)]{NIST:DLMF}.}.
The result of this first round is to extract 9,919 formulae from the DLMF
(see the F1 entries in Table \ref{firsttable1} for a chapter specific description of this).

{\small
\begin{table}[t]
\caption{Summary of DLMF chapters (Version 1.0.13, 2016-09-16, see Section \ref{firstextraction})
with corresponding 2-letter codes (2C), chapters numbers (C{\tt\#}), 
chapter names of the DLMF chapters, chapter formulas extracted from the first 
extraction scan (F1), and second extraction scan (F2).\\[-0.15cm]}
\begin{minipage}{0.51\linewidth}
\centering
%\vspace{0.2cm}
\begin{tabular}{ | l | c | l | c | c | }
\hline
\TT\TB {\bf 2C} & {\bf C\verb|#|} & {\bf Chapter Name} & {\bf F1} & {\bf F2} \\\hline
\TT\TB \verb|AL| & 1 & Algebraic and Analytic Methods & 624 & 85 \\\hline
\TT\TB \verb|AS| & 2 & Asymptotic Approximations & 375 & 56 \\\hline
\TT\TB \verb|NM| & 3 & Numerical Methods & 428 & 41 \\\hline
\TT\TB \verb|EF| & 4 & Elementary Functions & 529 & 466 \\\hline
\TT\TB \verb|GA| & 5 & Gamma Function & 181 & 56 \\\hline
\TT\TB \verb|EX| & 6 & Exponential, Logarithmic, Sine, & 110 & 44 \\
& & and Cosine Integrals & & \\\hline
\TT\TB \verb|ER| & 7 & Error Functions, Dawson's and & 158 & 80\\[0.0cm]
& & Fresnel Integrals & & \\\hline
\TT\TB \verb|IG| & 8 & Incomplete Gamma and Related & 272 & 120 \\
& & Functions & & \\\hline
\TT\TB \verb|AI| & 9 & Airy and Related Functions & 250 & 160 \\\hline
\TT\TB \verb|BS| & 10 & Bessel Functions & 904 & 355 \\\hline
\TT\TB \verb|ST| & 11 & Struve and Related Functions & 168 & 83 \\\hline
\TT\TB \verb|PC| & 12 & Parabolic Cylinder Functions & 205 & 100 \\\hline
\TT\TB \verb|CH| & 13 & Confluent Hypergeometric Functions\!  & 391 & 174 \\\hline
\TT\TB \verb|LE| & 14 & Legendre and Related Functions & 294 & 173 \\\hline
\TT\TB \verb|HY| & 15 & Hypergeometric Function & 206 & 174 \\\hline
\TT\TB \verb|GH| & 16 & Generalized Hypergeometric & 107 & 32 \\
& & Functions \& Meijer $G$-Function & & \\\hline
\TT\TB \verb|QH| & 17 & $q$-Hypergeometric and Related & 185 & 96 \\
& & Functions & & \\\hline
\end{tabular}
\end{minipage}
\begin{minipage}{0.5\linewidth}
\vspace{-0.3cm}
\centering
\begin{tabular}{ | l | c | l | c | c | }
\hline
\TT\TB {\bf 2C} & {\bf C\verb|#|} & {\bf Chapter Name} & {\bf F1} & {\bf F2}
\\\hline
\TT\TB \verb|OP| & 18 & Orthogonal Polynomials & 590 & 253 \\\hline
\TT\TB \verb|EL| & 19 & Elliptic Integrals & 674 & 391 \\\hline
\TT\TB \verb|TH| & 20 & Theta Functions & 115 & 84 \\\hline
\TT\TB \verb|MT| & 21 & Multidimensional Theta Functions\phantom{\,} & 62 & 7 \\\hline
\TT\TB \verb|JA| & 22 & Jacobian Elliptic Functions & 273 & 181 \\\hline
\TT\TB \verb|WE| & 23 & Weierstrass Elliptic and Modular & 199 & 70 \\
& & Functions & & \\\hline
\TT\TB \verb|BP| & 24 & Bernoulli and Euler Polynomials & 202 & 60 \\\hline
\TT\TB \verb|ZE| & 25 & Zeta and Related Functions & 176 & 50 \\\hline
\TT\TB \verb|CM| & 26 & Combinatorial Analysis & 215 & 54 \\\hline
\TT\TB \verb|NT| & 27 & Functions of Number Theory & 132 & 24 \\\hline
\TT\TB \verb|MA| & 28 & Mathieu Functions and Hill's & 442 & 186 \\
& & Equation  & & \\\hline
\TT\TB \verb|LA| & 29 & Lam\'{e} Functions & 251 & 72 \\\hline
\TT\TB \verb|SW| & 30 & Spheroidal Wave Functions & 161 & 46 \\\hline
\TT\TB \verb|HE| & 31 & Heun Functions & 177 & 33 \\\hline
\TT\TB \verb|PT| & 32 & Painlev\'{e} Transcendents & 342 & 59 \\\hline
\TT\TB \verb|CW| & 33 & Coulomb Functions & 210 & 92 \\\hline
\TT\TB \verb|TJ| & 34 & $3j$, $6j$, $9j$ Symbols &  68 & 30 \\\hline
\TT\TB \verb|FM| & 35 & Functions of Matrix Argument &  62 & 1 \\\hline
\TT\TB \verb|IC| & 36 & Integrals with Coalescing Saddles & 181 & 90 \\\hline
\end{tabular}
\end{minipage}
\label{firsttable1}
\end{table}
}

\subsection{Second extraction scan}
\label{secondextractionscansection}

In a second round of post processing, a subset of formulae are culled from the first
extraction scan (described in the previous section).
The formulae which are removed are those which contain the following \LaTeX\ commands
\verb|\sum|,
\verb|\int|,
\verb|\prod|,
\verb|\lim|,
\verb|\dots| (including all variants),
\verb|\sim|,
or the semantic macros
\verb|\BigO|,
\verb|\littleo|,
\verb|\fDiff|,
\verb|\bDiff|,
\verb|\cDiff|,
\verb|\asymp|,
or the environmnents
\verb|{cases}|,
\verb|{array}|,
\verb|{bmatrix}|,
\verb|{vmatrix}|,
\verb|{Bmatrix}|,
\verb|{pmatrix}|,
\verb|{Matrix}|,
\verb|{Lattice}|.
This is because either our CAS translator is unable to translate these effectively, or
the CAS is unable to verify mathematical objects such as these.
See Section \ref{OngoingSemanticenhancement} below for ongoing strategies for handling
formulae which fall into these categories.  Once these strategies are fully implemented,
then we will no longer need to cull formulae such as these. 

We have also utilized a new semantically enhanced macro \verb|\Wron| for 72, two-argument Wronskian
relations \cite[(1.13.4)]{NIST:DLMF}, so that the variable which is differentiated
against is precisely specified in an updated macro call.
In fact, we have continued to develop many new semantically enhanced \LaTeX\ macros (such as 
\verb|\Wron|) which are not in use in the DLMF, but are in use for the NIST Digital Repository
of Mathematical Formulae (see \cite{CohlDRMF2} for more details about this).
(e.g., \verb|\newcommand{\etpipm}[1]{\expe^{\pm2 \pi i/#1}}| which occur in 65 separate lines).
We also split equations with multiple equal \verb|=| signs (e.g., $a=b=c$ is split into two
formulae $a=b$ and $a=c$).
Furthermore, we split environments which contain \verb|\pm| (plus or minus) and \verb|\mp| (minus or plus)
commands into two separate formulae, each with their correct sign.
We also remove mathematical expressions which do not contain any of the relation commands
\verb|=|, \verb|<|, \verb|>|, \verb|\ne|, \verb|\le|, \verb|\ge|, \verb|\to|, \verb|\equiv|,
since their non-existence implies that there is no logical component for the mathematical 
expression to be verified against.
The result of this second round is to extract 4,078 formulae from the DLMF
(see the F2 entries in Table \ref{firsttable1} for a chapter specific description of this).

\subsection{Ongoing semantic enhancement for second extraction scan.}
\label{OngoingSemanticenhancement}

As mentioned above, we are currently not able to convert \LaTeX\ commands such as
\verb|\sum|, \verb|\int|, \verb|\prod|, \verb|\lim|, which represent sums, integrals,
products and limits, respectively. In this case, one draws (presentation) subscripts and
superscripts in order to provide critical semantic information for these mathematical operators.
In this case, there is no certainty in the specificity of the mathematical operation.

Another notation in which semantic capture is challenged is with the prime (e.g., $f'$,
double prime (e.g., $f''$), triple prime (e.g., $f'''$) or superscript parenthetical
(e.g., $f^{(iv)}$) notations for differentiation a given number of times.
This is the same issue that was encountered previously with the Wronskian relation. In order
for a CAS to be able to translate expressions such as these, we must provide the variable that
one is differentiating with respect to.  This is an example where a human is often able to
read a formula, and by knowing the context, be able to surmise what the variable is that one
is differentiating with respect to. Sometimes there is more than one variable in an expression
that one is differentiating with respect to, and if one does not provide the variable, then a
translator is unable to disambiguate the expression.
It is interesting to note that we were able to
enhance the semantics of 74 Wronskian relations by rewriting the macro so that it included the variable
that derivatives are taken with respect to as a parameter. A similar semantic enhancement
is possible for another 284 formulae where
the potentially ambiguous prime
notation
`\hspace{0.03cm}\textquotesingle\hspace{0.03cm}'
is used for derivatives.

The powerful tool used by the DLMF, \LaTeXML, is often able to guess a syntax tree when the \LaTeX\
commands \verb|\sum|, \verb|\int|, \verb|\prod|, and \verb|\lim| are used.  This is due to, for
instance, the utilization of the semantic \LaTeX\ \verb|\diff| macro for integrals. This is quite
effective, since it makes available the variable that one is integrating with respect to. For instance,
one may mark up a definite integral such as 
\[
\int_{a}^b f(x,y) \mathrm{d}x.
\]
by using semantic \LaTeX\ as follows \verb|\int_{a}^{b} f(x,y) \diff{x}|, which provides
the variable of integration \verb|x|. This information is essential for disambiguation of the 
involved operation variable, and therefore for converting the definite integral to a CAS representation.
Such content is not always easily accessible.  For instance, one is still unable to easily guess the
variable of differentiation when prime (etc.) notations for derivatives are used.  One is able to
facilitate the extraction of semantic content in formulae which contain mathematical \LaTeX\ such
as described above, by making the semantic content more easily available.  We would like to see this
type of semantic content made more fully available for translators, and this is an ongoing effort.

%TODO: I don't understand the following sentence
It is possible to incrementally enhance mathematical semantic expressions through optimization
and broader usage and development of semantic \LaTeX\ macros.  Take for instance the formula with 
the label \verb|{eq:OP.CP.SV.LR.1a}| \cite[(18.8.21)]{NIST:DLMF},
given by \\[-0.6cm]
\begin{center}
\verb|\lim_{\beta\to\infty}\JacobiP{\alpha}{\beta}{n}@{1-(\ifrac{2x}\beta)}}|\\
\hspace{-0.7cm}\verb|=\LaguerreL[\alpha]{n}@{x}|.
\end{center}
\vspace{-0.2cm}
In this example, our semantic enhancement process is made more effective by developing and utilizing
a new mathematical semantic \LaTeX\ macro with separate parameters and arguments 
(before and after the \verb|@| sign respectively) which is described as
\[
\verb|\Lim{#1}{#2}@{#3}|:=\lim_{\verb|#1|\to\verb|#2|} \verb|#3|.
\]
Here the first parameter of the macro call \verb|#1| is the variable
the limit is taken over, the second parameter of the macro call \verb|#2| is the destination of the limit
for \verb|#1|, and the argument \verb|#3| is the expression that the limit is taken over.  Utilizing 
this new macro, the above formula is then written with enhanced semantics as: \\[-0.6cm]
\begin{center}
\verb|\Lim{\beta}{\infty}@{\JacobiP{\alpha}{\beta}{n}@{1-(\ifrac{2x}\beta)}}|\\
\hspace{-0.7cm}\verb|=\LaguerreL[\alpha]{n}@{x}|.
\end{center}
\vspace{-0.2cm}
This new formula will have identical presentation to the previous formula,
and the semantics are more easily extracted (using for instance, the POM tagger).  We have also developed
semantic macros for sums, products, definite (or indefinite) integrals, as well as for antiderivatives, namely
\[
\verb|\Sum{#1}{#2}{#3}@{#4}|:=\sum_{\verb|#1|=\verb|#2|}^{\verb|#3|}\verb|#4|, \qquad
\verb|\Prod{#1}{#2}{#3}@{#4}|:=\prod_{\verb|#1|=\verb|#2|}^{\verb|#3|}\verb|#4|,
\]
\[
\verb|\Int{#1}{#2}@{#3}{#4}|:=\int_{\verb|#1|}^{\verb|#2|} \verb|#4|\,{\mathrm d}\verb|#3|, \qquad
\verb|\Antider@{#1}{#2}|:=\int \verb|#2|\,{\mathrm d}\verb|#1|.
\]
The incorporation of these semantic macros into semantic \LaTeX\ facilitates the translation of
the mathematical \LaTeX\ source expression to a CAS representation. It therefore improves one's ability
to automate injection of the mathematical content into CAS.

There often exists alternate usage of mathematical notations. For example, one often writes for a sum,
\[
\sum_{n\in A} f(n),
\]
where $A$ is some subset of the integers. It is not difficult to customize the macro definitions to be
capable of dealing with such notations, assuming that one has a precise description of $A$.

%\medskip
%\noindent {\sc We will have more information about the outcome for this process when the camera ready version is due. }

\section{CAS verification procedure for DLMF formulae}

We translate the 
semantic \LaTeX\ for our test dataset of DLMF formulae 
to \Maple\ CAS representation, using the 
tool described in \cite{CohlDRMF3}.
We use easily configurable settings to control our verifications.
Using the configuration files, we can control and customize settings
for the verification process, e.g., if one would like to perform numerical
evaluations on more complicated expressions involving differences or divisions of the left-hand sides and right-hand sides of the DLMF formulae.

\vspace{-0.30cm}
\subsection{Symbolic Verification of DLMF formulae}\label{sec:symbolic}
\vspace{-0.0cm}

Originally, we used the standalone \Maple{} \verb|simplify| function directly, to symbolically simplify
translated formulae.  See 
\cite{Koepf1991,Ballarin1995,Harrison1998,Baileyetal2014} for other examples of 
where \Maple{}\ and other CAS simplification procedures has been used elsewhere in 
the literature.
Symbolic simplification is performed either on the difference or the division
of the left-hand sides and the right-hand sides of extracted formulae.  Thus the expected outcome
should be respectively either a \verb|0| or \verb|1|.  Note that other outcomes, such as other numerical
outcomes, are particularly interesting, since these may be an indication of errors in the formulae.

In \Maple{}, symbolic simplifications are made using internally stored relations to other functions.
If a simplification is available, then in practice it often has to be performed over multiple
defined relevant relations. Often, this process fails and \Maple{} is unable to simplify the said
expression.  We have adopted some techniques which assist \Maple{} in this process.  For example, 
forcing an expression to be converted into another specific representation, in a pre-processing step,
could potentially improve the odds that \Maple{} is able to recognize a possible simplification.  By
trial-and-error, we discovered (and implemented) the following pre-processing steps which significantly
improve the simplification process:\\[-0.60cm]
\begin{itemize}
\item conversion to exponential representation;\\[-0.35cm]
\item conversion to hypergeometric representation;\\[-0.35cm]
\item expansion of expressions (for example \verb|(x+y)^2|); and\\[-0.35cm]
\item combined expansion and conversion processes. \\[-0.5cm]
\end{itemize}

\subsection{Constraint Handling.}
\label{ConstraintHandling}

Correct assumptions about variable domains are essential for CAS systems, and not 
surprisingly lead
to significant improvements in the CAS ability to simplify.  
The DLMF provides constraint (variable domain)
metadata for formulae, and as mentioned in Section \ref{firstextraction}, we have extracted this formula
metadata.  We have incorporated these constraints as assumptions for the simplification process.
Note however, that a direct translation of the constraint metadata is usually not sufficient for a CAS
to be able to understand it.
Furthermore, testing invalid values for numerical tests returns incorrect results.

For instance different symbols must be interpreted differently depending on the usage.
One must be able to interpret correctly certain notations of this kind. For instance, one must be
able to interpret the command \verb|a,b\in A|, which indicates that both variables
\verb|a| and \verb|b| are elements of the set \verb|A| (or more generally \verb|a_1,\dots,a_n\in A|).
Similar conventions are often used for variables being elements of other sets such as the sets of
rational, real or complex numbers, or for subsets of those sets.

Also, one must be able to interpret the constraints as variables in sets defined using an equals
notation such as \verb|n=0,1,2,\dots|, which indicates that the variable
\verb|n| is a integer greater than or equal to zero, or together \verb|n,m=0,1,2,\dots|, both the
variables \verb|n| and \verb|m| are elements of this set. Since mathematicians who write \LaTeX\ are
often casual about expressions such as these, one should know that \verb|0,1,2,\dots| is the same
as \verb|0,1,\dots|. Consistently, one must also be able to correctly interpret infinite sets (represented as
strings) such as
\verb|=1,2,\dots|,
\verb|=1,2,3,\dots|,
\verb|=-1,0,1,2,\dots|,
\verb|=0,2,4,\dots|, or even
\verb|=3,7,11,\dots|, or
\verb|=5,9,13,\dots|.
One must be able to interpret finite sets such as
\verb|=1,2|,
\verb|=1,2,3|, or
\verb|=1,2,\dots,N|.

An entire language of translation of mathematical notation must be understood in order for CAS to be able to understand constraints.
In mathematics, the syntax of constraints is often very compact and contains 
textual explanations.  Translating constraints from \LaTeX\ to CAS is a compact 
task because CAS only allow precise and strict syntax formats.  For example, 
the typical constraint $0 < x < 1$ is invalid if directly translated to \Maple{},
because it would need to be translated to two separate constraints, 
namely $x > 0$ and $x < 1$.  

We have improved the handling and translation of variable constraints/assumptions 
for simplification and numerical evaluation.  Adding assumptions about the 
constrained variables improves the effectiveness of \Maple{}’s \verb|simplify| 
function.
Our previous approach for constraint handling for numerical tests was to extract 
a pre-defined set of test values and to filter invalid values according to the 
constraints.  Because of this strategy, there often was no longer any valid values
remaining after the filtering. To overcome this issue, instead, we chose a single 
numerical value for a variable that appears in a pre-defined constraint.  For 
example, if a test case contains the constraint $0 < x < 1$, we chose
$x=\tfrac{1}{2}$.

A naive approach for this strategy, is to apply regular expressions to 
identify a match between a constraint and a rule.  However, we believed that this 
approach does not scale well when it comes to more and more pre-defined rules 
and more complex constraints.  Hence, we used the POM-tagger to create blueprints 
of the parse trees for pre-defined rules. For the example \LaTeX\ constraint 
\verb|$0 < x < 1$|, rendered as $0<x<1$, our 
textual rule is given by \\[-0.6cm]
\begin{center}
\verb|0 < var < 1 ==> 1/2|.
\end{center}
\vspace{-0.20cm}
The parse tree for this blueprint constraint contains five tokens, where \verb|var|
is an alphanumerical token that is considered to be a placeholder for a variable.

We can also distinguish multiple variables by adding an index to the placeholder.
For example, the rule we generated for the mathematical \LaTeX\ constraint 
\verb|$x,y \in \Real$|, where \verb|\Real| is the semantic macro which represents
the set of real numbers, and rendered as $x,y\in{\mathbb R}$,
is given by \\[-0.6cm]
\begin{center}
\verb|var1, var2 \in \Real ==> 3/2,3/2|.
\end{center}
\vspace{-0.20cm}
A constraint will match one 
of the blueprints if the number, the ordering, and the type of the tokens are equal.
Allowed matching tokens for the variable placeholders are Latin or Greek letters 
and alphanumerical tokens.

%\medskip
%\noindent {\sc We will have more information about the outcome for this process when the camera ready version is due. }

\vspace{-0.30cm}
\subsection{Numerical Verification of DLMF formulae}\label{sec:numerical}
\vspace{-0.0cm}

Due to the fact that CAS simplification verification is not extremely effective, we
also used our CAS translation to perform numerical testing as well.  To perform
automated numerical evaluations, we extracted all variables from the expression.
Variables are extracted by identifying all \verb|names| \cite{MPG}\footnote{A \textit{name} 
in \Maple{} is a sequence of one or more characters that uniquely identifies a command,
file, variable, or other entity.} from an expression.
This will also extract constants which need to be deleted from the list first.
Afterwards, we set each variable to a specific numerical value and numerically evaluate
the expression.  One is given the capability to choose the numerical values of the
given variables as either complex numbers, real numbers, or even as integers.
Given this typing of variables, we also check the values to ensure that they
are subject to the the constraints (see Section \ref{ConstraintHandling} above).
We allow for the definition of a set of numerical values for variables that we want
to verify, and for multiple variable choices, the power-set of these choices is
looped over (we evaluate all combinations of variable-value pairs).  We consider an evaluation to be successful if the outcome is below a given threshold (currently set at \verb|0.001|) compared with a certain precision (currently set at \verb|10|).

%\medskip
%\noindent {\sc We will have more information about the outcome for this process when the camera ready version is due. }

\vspace{-0.30cm}
\section{Summary}\label{sec:summary}
\vspace{-0.0cm}

We have created a test dataset\footnote{This dataset is available on request to the authors.} of 4,078 
semantic \LaTeX\ formulae, extracted from
the DLMF.  We translated each test case to a representation in \Maple{}
and used \Maple's {\tt simplify} function on the formula difference to verify that the
translated formulae remain valid. Our forward translation tool (Section~\ref{sec:LatexToCAS})
was able to translate 2,405 (approx.~59\%) test cases. 
Most likely, a translation failed because it encountered a DLMF/DRMF semantic macro without a known translation to \Maple{} (1,021 of non-translated cases, approx.~61\%).
An example of such a non-supported macros is the (basic) $q$ -hypergeometric function~\cite[(17.4.1)]{NIST:DLMF}.
In other cases, translation of expressions failed because they contained insufficient semantic information, such as when the prime and superscript notations for derivatives are used (Section \ref{OngoingSemanticenhancement}), or encountered assorted errors in the translation engine. 
Such assorted errors include unimplemented grammar mappings from the POM tagger such as handling subsuperscript outputs or overlining expressions for complex conjugations. 

We applied our symbolic verification techniques to the 2,405 translated expressions.
The proposed simplification was able to verify 481 of these expressions (20\%). 
Pre-conversion improved the effectiveness of {\tt simplify} and was used to
convert the translated expression to a different form before simplification of the formula
difference. 
We used conversions to exponential and hypergeometric form and expanded the
translated expression. 
Those pre-processed manipulations increased the number of formulae verified from 481 to 660 (approx.~27.4\%).
The remaining 1,745 test cases were translated but not verified.
Note that verification should also fail because either expression would not contain a logical relation and therefore are unverifiable (Section \ref{secondextractionscansection}), but these are filtered out in a pre-processing step, or that the CAS is not yet sophisticated enough to verify (such as in the case of asymptotic representations). 

We have also applied several numerical tests to these remaining test cases (Section~\ref{sec:numerical}).
The results of the numerical tests strongly depend on the tested values. 
Besides the defined numerical values for general tests, we applied special values for certain constraints.
This was realized by our new approach of applying rules for blueprints of common constraints (Section~\ref{ConstraintHandling}).
In automated evaluations, we performed numerical tests by setting variables in test expressions to values contained in the set $\{-0.5, 0.5, 1.5\}$.
In the cases where the tested expression is an equation, we tested the difference between the left- and the right-hand sides of the equations.
For other relations, we tested if the relation still remains for the calculated numerical values.
With this set of the three values and the set of special values depending on additional constraints, 418 of the remaining unverified test cases return a valid output (approx.~24\%).
14 numerical tests stopped, because they exceeded our pre-defined time-limit of 300 seconds.
In 892 cases (approx.~51.1\%), the calculated values were above our given threshold of \verb|0.001|. Note that numerical verifications may fail because of one of four reasons:
\begin{enumerate}
\item\label{reasons:i} the numerical engine tests invalid combinations of values;
\item\label{reasons:ii} the translation is incorrect;
\item\label{reasons:iii} there may be an error in the DLMF source; or
\item\label{reasons:iv} there may be an error in \Maple.
\end{enumerate}
Typical originations for reason~\ref{reasons:i} are errors produced in the translations of constraints (Section \ref{ConstraintHandling}), and may prevent the numerical test engine from using valid numerical values for evaluation. 
Another typical problem is missing information from  context, such as if a variable was pre-defined by substitution, but the test case misses this information. 
For example, \cite[(9.6.1)]{NIST:DLMF} defines $\zeta = \frac{2}{3}z^{3/2}$. 
The following equations are only valid considering the pre-defined numerical value for $\zeta$. 
Furthermore, this equation is a definition of $\zeta$. 
If we test this equation by simplifications or numerical evaluations, we get invalid results. 
We call such failed numerical evaluations by reason~\ref{reasons:i}, \textit{false positives}, since they were marked as particularly interesting cases (for reasons~\ref{reasons:ii}-\ref{reasons:iv}) but they were caused by missing semantic information originating from the source. 
It seems that the number of false positive results is very high and plan to reduce this in the future. 
Table~\ref{tab:results} gives an overview of the symbolic and numerical testing for each DLMF chapter. 
The overview of translations (T) per chapter reveals the weakness of our translation engine especially in the Chapters 17 (\verb|QH|) and 34 (\verb|TJ|). 
A high ratio of symbolically verified and translated formulae indicates a large success rate for translating macros into \Maple{} functions. 
This is one explanation for the fact that our best result comes from Chapter 4 \verb|EF|. 

{\small
\begin{table}[t]
\caption{Overview of symbolic and numerical testing for each DLMF chapter with the corresponding DLMF chapter 2-letter codes (2C), chapters numbers (C{\tt\#}), number of extracted expressions of the second extraction scan (F2), number of translated expressions (T) with approximated percentages, number of translated and symbolically verified cases (TV\textsubscript{s}) with approximated percentages, and number of translated cases where the numerical evaluations returned valid outputs for the set of test values $\{-0.5, 0.5, 1.5\}$ (TV\textsubscript{n}) with approximated percentages (excluding the number of symbolically verified translations). The best results of each technique are highlighted in bold.\\[-0.15cm]}
\begin{minipage}{0.51\linewidth}
\centering
%\vspace{0.2cm}
\begin{tabular}{ | c | c | r | r r | r r | r r | }
\hline
\TT\TB {\bf 2C} & {\bf C\verb|#|} & {\bf F2} & \multicolumn{2}{|c|}{\bf T} & \multicolumn{2}{|c|}{\bf TV\textsubscript{s}} & \multicolumn{2}{|c|}{\bf TV\textsubscript{n}} \\\hline
\TT\TB \verb|AL| & 1 & 85 	& 44 & (51.8\%) & 4 	& (4.7\%) 	& 8  & (9.9\%) \\\hline
\TT\TB \verb|AS| & 2 & 56 	& 27 & (48.2\%) & 2 	& (3.6\%) 	& 0  & (0\%) \\\hline
\TT\TB \verb|NM| & 3 & 41 	& 34 & (82.9\%) & 1 	& (2.4\%) 	& 1  & (2.5\%) \\\hline
\TT\TB \verb|EF| & 4 & 466 	& 406& (87.1\%) & {\bf182} 	& {\bf(39.1\%)} 	& 61 & (21.5\%) \\\hline
\TT\TB \verb|GA| & 5 & 56 	& 39 & (69.6\%) & 10 	& (17.9\%) 	& {\bf15} & {\bf(32.6\%)} \\\hline
\TT\TB \verb|EX| & 6 & 44 	& 13 & (29.5\%) & 0 	& (0\%) 	& 1  & (2.3\%) \\\hline
\TT\TB \verb|ER| & 7 & 80 	& 60 & (75.0\%)   & 24 	& (30\%) 	& 11 & (19.6\%) \\\hline
\TT\TB \verb|IG| & 8 & 120 	& 89 & (74.2\%) & 24 	& (20\%) 	& 13 & (13.5\%) \\\hline
\TT\TB \verb|AI| & 9 & 160 	& 71 & (44.4\%) & 8 	& (5\%) 	& 19 & (12.5\%) \\\hline
\TT\TB \verb|BS| & 10 & 355 & 174& (49.0\%)   & 44 	& (12.4\%) 	& 47 & (11.9\%) \\\hline
\TT\TB \verb|ST| & 11 & 83 	& 71 & (85.5\%) & 20 	& (24.1\%) 	& 11 & (17.5\%) \\\hline
\TT\TB \verb|PC| & 12 & 100 & 71 & (71.0\%)   & 14 	& (14\%) 	& 12 & (14\%) \\\hline
\TT\TB \verb|CH| & 13 & 174 & 168& (96.6\%) & 61 	& (35.1\%)	& 24 & (21.2\%) \\\hline
\TT\TB \verb|LE| & 14 & 173 & 163& (94.2\%) & 31 	& (17.9\%) 	& 37 & (26.1\%) \\\hline
\TT\TB \verb|HY| & 15 & 174 & {\bf170}& {\bf(97.7\%)} & 39 	& (22.4\%) 	& 34 & (25.2\%) \\\hline
\TT\TB \verb|GH| & 16 & 32 	& 18 & (56.3\%) & 3 	& (9.4\%) 	& 0  & (0\%)\\\hline
\TT\TB \verb|QH| & 17 & 96 	& 0  & (0\%) & \multicolumn{2}{|c|}{---} & \multicolumn{2}{|c|}{---} \\\hline
\TT\TB \verb|OP| & 18 & 253 & 141& (55.7\%) & 58 	& (22.9\%) 	& 23 & (11.8\%)\\\hline
 & & & & & & & & \\\hline
\end{tabular}
\end{minipage}
\begin{minipage}{0.5\linewidth}
%\vspace{-0.3cm}
\centering
\begin{tabular}{ | c | c | r | r r | r r | r r | }
\hline
\TT\TB {\bf 2C} & {\bf C\verb|#|} & {\bf F2} & \multicolumn{2}{|c|}{\bf T} & \multicolumn{2}{|c|}{\bf TV\textsubscript{s}} & \multicolumn{2}{|c|}{\bf TV\textsubscript{n}} \\\hline
\TT\TB \verb|EL| & 19 & 391 & 151& (38.6\%) & 31	& (7.9\%)	& 31 & (8.6\%) \\\hline
\TT\TB \verb|TH| & 20 & 84 	& 56 & (66.7\%) & 6 	& (7.1\%) 	& 0  & (0\%) \\\hline
\TT\TB \verb|MT| & 21 & 7 	& 2  & (28.6\%) & 0 	& (0\%) 	& 0  & (0\%) \\\hline
\TT\TB \verb|JA| & 22 & 181 & 146& (80.7\%) & 59 	& (32.6\%) 	& 34 & (27.9\%) \\\hline
\TT\TB \verb|WE| & 23 & 70 	& 28 & (40.0\%)   & 0 	& (0\%) 	& 0  & (0\%) \\\hline
\TT\TB \verb|BP| & 24 & 60 	& 34 & (56.7\%) & 7 	& (11.2\%) 	& 10 & (18.9\%) \\\hline
\TT\TB \verb|ZE| & 25 & 50 	& 27 & (54.0\%)  & 12 	& (24\%) 	& 6  & (15.8\%) \\\hline
\TT\TB \verb|CM| & 26 & 54 	& 27 & (50.0\%)  & 9 	& (16.7\%)	& 10 & (22.2\%) \\\hline
\TT\TB \verb|NT| & 27 & 24 	& 4  & (16.7\%) & 0 	& (0\%) 	& 0  & (0\%) \\\hline
\TT\TB \verb|MA| & 28 & 186 & 43 & (23.1\%) & 6 	& (3.2\%) 	& 6  & (3.3\%) \\\hline
\TT\TB \verb|LA| & 29 & 72 	& 17 & (23.6\%) & 0 	& (0\%) 	& 1  & (1.4\%) \\\hline
\TT\TB \verb|SW| & 30 & 46 	& 11 & (23.9\%) & 0 	& (0\%) 	& 0  & (0\%) \\\hline
\TT\TB \verb|HE| & 31 & 33 	& 28 & (84.8\%) & 4 	& (12.1\%) 	& 0  & (0\%) \\\hline
\TT\TB \verb|PT| & 32 & 59 	& 31 & (52.5\%) & 0 	& (0\%) 	& 1  & (1.7\%) \\\hline
\TT\TB \verb|CW| & 33 & 92 	& 18 & (19.6\%) & 1 	& (1.1\%) 	& 1  & (1.1\%) \\\hline
\TT\TB \verb|TJ| & 34 & 30 	& 0  & (0\%) & \multicolumn{2}{|c|}{---} & \multicolumn{2}{|c|}{---} \\\hline
\TT\TB \verb|FM| & 35 & 1 	& 0  & (0\%) & \multicolumn{2}{|c|}{---} & \multicolumn{2}{|c|}{---} \\\hline
\TT\TB \verb|IC| & 36 & 90 	& 23 & (25.6\%) & 0 	& (0\%)		& 0  & (0\%) \\\hline\hline
\TT\TB $\Sigma$ &  & 4087	& 2405 & (58.8\%) & 660 	& (16.1\%)		& 418  & (12.2\%) \\\hline
\end{tabular}
\end{minipage}
\label{tab:results}
\end{table}
}

Originally, the verification rules were developed to identify errors in the translation engine, see reason~\ref{reasons:ii}. 
However, further investigations of the 892 cases reveals a sign error in the DLMF (reason~\ref{reasons:iii}), and this corresponded to 
\cite[(14.5.14)]{NIST:DLMF}, namely
\[
{\mathsf Q}_\nu^{-1/2}(\cos\theta)=\left(\frac{\pi}{2\sin\theta}\right)^{1/2}
\frac{\cos((\nu+\frac12)\theta)}{\nu+\frac12},
\]
where ${\mathsf Q}_\nu^\mu$ is the Ferrers function of the second kind. 
This error can be found on \cite[p.~359]{NISTHandbook2010}, and was reported in DLMF Version 1.0.16 on September 18, 2017. Other cases where formulae seem to be unverified by numerical tests are for DLMF formulas which are valid on Riemann surfaces, but fail because \Maple{} uses specific choices of branch values (reason~\ref{reasons:iv}).
Using our constraint blueprints we were able to identify when a constraint does not
match the set of rules.  In fact, our constraint handling identified a missing comma 
after the -2 in the constraint of \cite[(10.16.7)]{NIST:DLMF}, originally
printed as $2\nu=-1, -2 -3, \ldots$. 
This error can be found on \cite[p.~228]{NISTHandbook2010}, and will be 
reported in the DLMF Version 1.0.19, which is scheduled to be published on June 15, 2018.
%A special case has proven, that even the varified translated test cases do not need to be correct.
%The equation \cite[(7.18.4)]{NIST:DLMF}
%\begin{equation}\label{eq:7.18.4}
%\frac{{\mathrm{d}}^{n}}{{\mathrm{d}z}^{n}}\left(e^{z^{2}}\operatorname{erfc}z%
%\right)=(-1)^{n}\,2^{n}\,n!\,e^{z^{2}}\mathop{\mathrm{i}^{n}\,\mathrm{erfc}}\left(z%
%\right),
%\end{equation}
%where $n \in \mathbb{N}_0$, $e$ is the base of the natural logarithm, and $\operatorname{erfc}$ is the complementary error function, is part of our test set. 
%While the previous and following equations were successfully verified by simplifications, they failed for~(\ref{eq:7.18.4}).
%Numerical verifications of~(\ref{eq:7.18.4}) exceeded time limitations.
%Further manual investigations show, that the left-hand side of~(\ref{eq:7.18.4}) was correctly translated to \verb|diff((exp((z)^(2))*erfc(z)), [z$(n))]|. 
%The {\tt simplify} function of \Maple{ 2016} falsely returns $0$ for the translated left-hand side.
We are planning to extend the blueprint approach to solve the more general 
problem of translating constraints to CAS.
We continue to update our checking procedure in search of more DLMF and \Maple{} errors/inconsistencies, and are constantly improving the translation engine.
\\[-0.3cm]

%We are currently preparing a test dataset of 100 formulae, sampled from the DLMF/DRMF. In the near
%future (prior to the camera-ready copy deadline), we will provide a comprehensive quantitative
%evaluation based on this dataset.\\[1.6cm]

%\vspace{0.10cm}
%\{\it section*{Acknowledgements}}
\noindent
\hspace{-0.111cm}
{\bf\large\!\hspace{0.04cm}Acknowledgements.\hspace{0.02cm}}
We are indebted to Bruce Miller and Abdou Youssef for valuable
discussions and for the development of the custom macro set of
semantic mathematical \LaTeX\ macros used in the DLMF, and for the
development of the POM tagger, respectively. We also thank the DLMF editors for their assistance and support.
%\\[-0.8cm]

%\bibliographystyle{alpha}
%\bibliography{sample}
%\bibliographystyle{unsrt}
%\bibliography{/home/hcohl/tex/refbib}
%\bibliography{refbib}

\printbibliography[notcategory=preprintref]

\end{document}